\documentclass[10pt,reqno,a4paper]{amsart}
\usepackage[british]{babel}
\usepackage{amssymb,amstext,amsthm,eucal,bbm,mathrsfs,amscd,bbold}
\usepackage[margin=1.5cm]{geometry}
\usepackage{array}
\usepackage[svgnames,table]{xcolor}
\usepackage[utf8]{inputenc}
\usepackage{enumitem}
\usepackage[small]{eulervm}
\usepackage{tgpagella}
\usepackage{tikz,pgfplots}
\usetikzlibrary{calc}
\pgfplotsset{compat=1.14}
\usepackage[unicode]{hyperref}
\hypersetup{%
  pdftitle   = {Classification of kinematical Lie algebras},
  pdfkeywords = {Lie algebra, deformations, kinematical},
  pdfauthor  = {José Figueroa-O'Farrill},
  pdfcreator = {\LaTeX\ with package \flqq hyperref\frqq},
  linkcolor=NavyBlue,
  citecolor=ForestGreen,
  urlcolor=OrangeRed,
  anchorcolor=OrangeRed,
  colorlinks=true
}
\theoremstyle{plain}

\theoremstyle{definition}

\renewcommand{\a}{\mathfrak{a}}

\newcommand{\g}{\mathfrak{g}}
\renewcommand{\k}{\mathfrak{k}}

\newcommand{\p}{\mathfrak{p}}
\renewcommand{\c}{\mathfrak{c}}
\newcommand{\co}{\mathfrak{co}}
\newcommand{\e}{\mathfrak{e}}
\newcommand{\n}{\mathfrak{n}}
\renewcommand{\r}{\mathfrak{r}}

\newcommand{\so}{\mathfrak{so}}

\newcommand{\s}{\mathfrak{s}}

\newcommand{\R}{\boldsymbol{R}}
\newcommand{\B}{\boldsymbol{B}}

\renewcommand{\P}{\boldsymbol{P}}

\newcommand{\RR}{\mathbb{R}}

\newcommand{\CC}{\mathbb{C}}

\newcommand{\Bbar}{\bar{\B}}
\newcommand{\Pbar}{\bar{\P}}
\definecolor{gris}{rgb}{0.8,0.8,0.8}

\allowdisplaybreaks[0]
\begin{document}

\title{Classification of kinematical Lie algebras}
\author{José M. Figueroa-O'Farrill}
\address{Maxwell Institute and School of Mathematics, The University
  of Edinburgh, James Clerk Maxwell Building, Peter Guthrie Tait Road,
  Edinburgh EH9 3FD, United Kingdom}
\begin{abstract}
  We summarise the classification of kinematical Lie algebras in
  arbitrary dimension and indicate which of the kinematical Lie
  algebras admit an invariant inner product.
\end{abstract}
\maketitle

\section{Introduction}
\label{sec:introduction}

By a \textbf{kinematical Lie algebra} in dimension $D$, we mean a
real $\tfrac12 (D+1)(D+2)$-dimensional Lie algebra with generators
$R_{ab} = - R_{ba}$, with $1\leq a,b \leq D$, spanning a Lie
subalgebra $\r \cong \so(D)$; that is,
\begin{equation}
  \label{eq:so}
  [R_{ab}, R_{cd}] = \delta_{bc} R_{ad} -  \delta_{ac} R_{bd} -
  \delta_{bd} R_{ac} +  \delta_{ad} R_{bc},
\end{equation}
and $2D+1$ generators $B_a$, $P_a$ and $H$ which transform according
to the vector, vector and scalar representations of $\so(D)$,
respectively -- namely,
\begin{equation}
  \label{eq:kin}
    [R_{ab}, B_c]= \delta_{bc} B_a - \delta_{ac} B_b\qquad
    [R_{ab}, P_c] = \delta_{bc} P_a - \delta_{ac} P_b\qquad\text{and}\qquad
    [R_{ab}, H] = 0.
\end{equation}
The rest of the brackets between $B_a$, $P_a$ and $H$ are only subject
to the Jacobi identity: in particular, they must be $\r$-equivariant.
The kinematical Lie algebra where those additional Lie brackets vanish
is called the \textbf{static} kinematical Lie algebra and shall be
denoted $\s$.  Every other kinematical Lie algebra will be, by
definition, a deformation of $\s$.  A partial converse, which is an
easy consequence of the Hochschild--Serre factorisation theorem
\cite{MR0054581}, is that for $D\geq 3$ every deformation of $\s$ is
kinematical.  This fails for $D=2$ because $\so(2)$ is not semisimple
and there are many more deformations of $\s$ in $D=2$ than the ones
which concern us in this paper.

Up to isomorphism, there is only one kinematical Lie algebra in $D=0$:
it is one-dimensional and hence abelian. For $D=1$, $\r = 0$ and hence
any three-dimensional Lie algebra is kinematical.  The classification
is therefore the same as the celebrated Bianchi classification of
three-dimensional real Lie algebras \cite{Bianchi}.  Chronologically,
the next classification was for $D=3$ by Bacry and Nuyts
\cite{MR857383}, following up from earlier work of Bacry and
Lévy-Leblond \cite{MR0238545}. A deformation theory approach to that
(and related) classifications is given in \cite{JMFKinematical3D}
based on earlier work \cite{JMFGalilean}. This same approach has been
used in \cite{JMFKinematicalHD} to classify kinematical Lie algebras
for $D\geq 4$ and in \cite{TAJMFKinematical2D} for $D=2$. The purpose
of this brief note is to summarise the results of the papers
\cite{JMFKinematical3D, TAJMFKinematical2D, JMFKinematicalHD}, which
contain the details of the necessary calculations.

Recall that a Lie algebra $\k$ is said to be \textbf{metric} if it admits
an ad-invariant (also called \emph{associative}) inner product; that
is, a non-degenerate symmetric bilinear form $\left<-,-\right>: \k \times \k \to
\RR$ which satisfies
\begin{equation*}
  \left<[x,y],z\right> = \left<x,[y,z]\right>\qquad\forall x,y,z \in \k.
\end{equation*}
It follows from Cartan's semisimplicity criterion that semisimple Lie
algebras are metric relative to the Killing form, but there are
non-semisimple metric Lie algebras, where $\left<-,-\right>$ is an
additional piece of data.  For example, any inner product on an
abelian Lie algebra is invariant.  We also indicate in our results
which of the Lie algebras in our classifications are metric.

\subsection*{Notation}
\label{sec:notation}

We use the perhaps non-standard notation for Lie algebras described in
Table~\ref{tab:notation}.  A caret adorning a symbol means a
nontrivial central extension, e.g., $\hat\g$ is the Bargmann algebra,
et cetera.

\begin{table}[h!]
  \centering
  \caption{Notation for Lie algebras}
  \label{tab:notation}
  \begin{tabular}{>{$}c<{$}|l}
    \multicolumn{1}{c|}{Notation} & \multicolumn{1}{c}{Name}\\\hline
    \a & abelian\\
    \s & static\\
    \n_+ & (euclidean) Newton\\
    \n_- & (lorentzian) Newton\\
    \e & euclidean\\
  \end{tabular}
  \hspace{3cm}
  \begin{tabular}{>{$}c<{$}|l}
    \multicolumn{1}{c|}{Notation} & \multicolumn{1}{c}{Name}\\\hline
    \p & Poincaré\\
    \so & orthogonal\\
    \co & orthogonal + dilatation\\
    \g & galilean\\
    \c & Carroll\\
  \end{tabular}
\end{table}

\section{$D=1$: Bianchi revisited}
\label{sec:bianchi}

Apart from the abelian Lie algebra (Bianchi~I) and the simple
three-dimensional Lie algebras (Bianchi~VIII and IX), all other
three-dimensional Lie algebras have the structure of an abelian
two-dimensional Lie algebra extended by an outer derivation.  Letting
$B$ and $P$ denote the generators of the abelian Lie algebra and $H$
the outer derivation, we arrive at the following nonzero brackets:
\begin{equation}
  [H,B] = a B + c P \qquad\text{and}\qquad [H,P] = b B + d P,
\end{equation}
which can be brought to a normal form.  Table~\ref{tab:summaryD1} lists
the different isomorphism classes and relates them to the Bianchi
classification.  It should be mentioned that the parameter $\gamma$ in
Bianchi VI is not the traditional parameter, but a
``compactification'' to the interval.

\begin{table}[h!]
  \centering
  \caption{Kinematical Lie algebras in $D=1$}
  \label{tab:summaryD1}
  \rowcolors{2}{blue!10}{white}
  \begin{tabular}{l|*{3}{>{$}l<{$}}|l|c}
    \multicolumn{1}{c|}{Bianchi} & \multicolumn{3}{c|}{Nonzero Lie brackets} & \multicolumn{1}{c|}{Comments} & Metric?\\\hline
    I & & & & $\a$ ($\cong\s$) & \checkmark \\
    II & [H,B] = P & & & $\g$ ($\cong\c$)& \\
    III & & [H,P] = P & & & \\    
    IV & [H,B] = B + P & [H, P] = P & & & \\
    V & [H,B] = B & [H,P] = P & & & \\
    VI$_0$ & [H,B] = -B & [H,P] = P & & $\n_-$ ($\cong\p$) & \\
    VI$_\gamma$ & [H,B] = \gamma B & [H,P] = P & & $0 \neq \gamma \in (-1,1)$ & \\
    VII$_0$ & [H,B] = P & [H,P] = - B & & $\n_+$ ($\cong\e$) & \\
    VII$_\alpha$ & [H,B] = \alpha B + P & [H,P] = \alpha P - B & & $\alpha > 0$ & \\
    VIII &  [H,B] = P & [H,P] = -B & [B,P] = - H & $\so(1,2)$ & \checkmark\\
    IX &  [H,B] = P & [H,P] = -B & [B,P] = H & $\so(3)$ & \checkmark\\
  \end{tabular}
\end{table}

In this dimension we already see many of the types of kinematical Lie
algebras which exist for generic $D$.  There are some isomorphisms
which are low-dimensional accidents, such as between the Carroll and
galilean algebras, between the Newton and euclidean/Poincaré
algebras and also between the de~Sitter/hyperbolic and anti~de~Sitter
algebras.  Of the Lie algebras in Table~\ref{tab:summaryD1}, only the
abelian (I) and simple (VIII, IX) cases are metric.

\section{$D=2$}
\label{sec:d=2}

This and the next dimension have a richer set of kinematical Lie
algebras than for generic $D$.  In the case of $D=2$ it has to do with
the $\so(2)$-invariant symplectic structure on the vector
representation.  In this case the rotational algebra is
one-dimensional and hence abelian and equation \eqref{eq:kin} takes
the simpler form
\begin{equation}
  \label{eq:static-R}
  [R,B_a] = \epsilon_{ab} B_b \qquad\text{and}\qquad
  [R,P_a] = \epsilon_{ab} P_b,
\end{equation}
with $\epsilon_{ab}$ the Levi-Civita symbol normalised to
$\epsilon_{12} = +1$.  We may diagonalise the action of $R$ by complexifying. To this end we
introduce $\B = B_1 + i B_2$ and $\P = P_1 + i P_2$ and extend the Lie
brackets complex-linearly, so that now
\begin{equation}
  \label{eq:static-C}
  [R,\B] = - i \B \qquad\text{and}\qquad [R, \P] = -i \P.
\end{equation}
We also have $\Bbar = B_1 - i B_2$ and $\Pbar = P_1 - i P_2$, which
satisfy
\begin{equation}
  [R,\Bbar] = i \Bbar \qquad\text{and}\qquad [R,\Pbar] = i \Pbar.
\end{equation}
The \emph{complex} span of $R,H,\B,\P,\Bbar,\Pbar$ subject to the
brackets~\eqref{eq:static-C} (and their complex conjugates) defines a
complex Lie algebra $\s_\CC$.  This complex Lie algebra has a
conjugation (that is, a complex-antilinear involutive automorphism)
denoted by $\star$ and defined by $H^\star = H$, $R^\star = R$,
$\B^\star = \Bbar$ and $\P^\star = \Pbar$.  We see that the real Lie
subalgebra of $\s_\CC$ consisting of real elements (i.e., those $X \in
\s_\CC$ such that $X^\star = X$) is the static kinematical Lie algebra
$\s$.  The same holds for any other kinematical Lie algebra in $D=2$:
its complexification admits the above conjugation.  We find it
convenient in the summary given in Table~\ref{tab:summaryD2} to use
the complex form of the Lie algebra.

The kinematical Lie algebras below the horizontal line in
Table~\ref{tab:summaryD2} are unique to $D=2$ and owe their existence
to the invariant symplectic structure $\epsilon_{ab}$ or,
equivalently, to the complex structure which in the complex version of
the algebra is simply multiplication by $i$.  We see that in $D=2$,
the Carroll, euclidean and Poincaré algebras are metric, as well as
two of the kinematical Lie algebras which are unique to this dimension.

\begin{table}[h!]
  \centering
  \setlength{\extrarowheight}{2pt}  
  \caption{Kinematical Lie algebras in $D=2$ (complex form)}
  \label{tab:summaryD2}
  \rowcolors{2}{blue!10}{white}
  \begin{tabular}{*{5}{>{$}l<{$}}|l|c}
    \multicolumn{5}{c|}{Nonzero Lie brackets} & \multicolumn{1}{c|}{Comments} & Metric?\\\hline
    \relax & & & & & $\s$\\
    \relax [H,\B] = \P & & & & & $\g$& \\
    \relax [H,\B] = \B & [H,\P] = -\P & & & & $\n_-$& \\
    \relax [H,\B] = i \B & & & & & $\n_+$ & \\
    \relax [H,\B] = \B & [H,\P] = (\lambda + i \theta) \P & & & & $\lambda \in (-1,1]$, $\theta \in \RR$& \\
    \relax [H,\B] = \B & [H,\P] = \B + \P & & & & & \\
    \relax & & & [\B,\Pbar] = H & & $\c$&  \checkmark\\
    \relax [H,\B] = \B & [H,\P] = -\P & & [\B,\Pbar] = 2(R - i H) & & $\so(3,1)$&  \checkmark\\
    \relax & [H,\P] = \B & & [\B,\Pbar] = -2 H & [\P,\Pbar] = -2i R & $\e$&  \checkmark\\
    \relax & [H,\P] = - \B & & [\B,\Pbar] = -2 H & [\P,\Pbar] = 2 i R & $\p$&  \checkmark\\    
    \relax [H,\B] = -\P & [H,\P] =  \B & [\B,\Bbar] = -2 i R & [\B,\Pbar]= -2 H & [\P,\Pbar] = -2 i R &  $\so(4)$&  \checkmark\\
    \relax [H,\B] = -\P & [H,\P] =  \B & [\B,\Bbar] =  2 i R & [\B,\Pbar]= 2 H & [\P,\Pbar] =  2 i R &  $\so(2,2)$&  \checkmark \\\hline
    \relax & & [\B, \Bbar]=i H & & [\P, \Pbar] = i H &&  \checkmark\\
    \relax [H,\B] = i \B & & [\B,\Bbar] = i H & & [\P,\Pbar] = i (H+R) & & \checkmark\\
    \relax & & [\B,\Bbar] = i H & & & & \\
    \relax [H,\B] = \P & & [\B,\Bbar] = i H & & & & \\
    \relax [H,\B] = \pm i \B & & [\B,\Bbar] = i H & & & & \\
  \end{tabular}
\end{table}

\section{$D=3$}
\label{sec:d=3}

This is the other classical case.  It is convenient here and in the
case of $D>3$ as well, to follow the abbreviated notation introduced
in \cite{MR0238545}, by which we let $\B$ and $\P$ stand for the
$D$-dimensional vectors of generators $B_a$ and $P_a$, and write the
Lie brackets without indices, with the understanding that the indices
in the LHS also appear on the RHS, albeit possibly contracted with the
rotationally invariant tensors: the Kronecker $\delta$ and the
Levi-Civita $\epsilon$.

All kinematical Lie algebras share the Lie brackets in
equations~\eqref{eq:so} and \eqref{eq:kin}, which in abbreviated
notation are written as
\begin{equation}
  [\R,\R] = \R \qquad [\R,\B] = \B \qquad [\R, H ] = 0
  \qquad\text{and}\qquad [\R, \P] = \P.
\end{equation}
The kinematical Lie algebras in Table~\ref{tab:summaryD3} which lie
below the line are unique to $D=3$: indeed, they owe their existence
to the vector product in $\RR^3$, which is invariant under rotations.
In $D=3$ the metric Lie algebras are the simple Lie algebras and in
addition four of the Lie algebras which are unique to this dimension.

\begin{table}[h!]
  \centering
  \caption{Kinematical Lie algebras in $D=3$}
  \label{tab:summaryD3}
  \rowcolors{2}{blue!10}{white}
  \begin{tabular}{*{5}{>{$}l<{$}}|l|c}
    \multicolumn{5}{c|}{Nonzero Lie brackets} & \multicolumn{1}{c|}{Comments} & Metric?\\\hline
    \relax & & & & & $\s$& \\
    \relax [H ,\B] = -\P & & & & & $\g$& \\
    \relax [H ,\B] = - \B & [H ,\P] = \P & & & & $\n_-$& \\
    \relax [H ,\B] = \P & [H ,\P] = - \B & & & & $\n_+$& \\
    \relax [H ,\B] = \gamma \B & [H ,\P] = \P & & & & $\gamma \in (-1,1)$& \\
    \relax [H ,\B] = \B & [H ,\P] = \P & & & & & \\
    \relax [H ,\B] = \alpha \B + \P & [H ,\P] = \alpha \P - \B & & & & $\alpha > 0$& \\
    \relax [H ,\B] = \B + \P & [H , \P] = \P & & & & & \\
    \relax & & [\B,\P] = H  & & & $\c$& \\
    \relax [H ,\B] = \P & & [\B,\P] = H  & [\B,\B] = \R & & $\e$& \\
    \relax [H ,\B] = -\P & & [\B,\P] = H  & [\B,\B] = -\R & & $\p$& \\
    \relax [H ,\B] = \B & [H ,\P] = -\P &  [\B,\P] = H  - \R & & & $\so(4,1)$&  \checkmark\\
    \relax [H ,\B] = \P & [H ,\P] = -\B & [\B,\P] = H  &  [\B,\B]= \R &  [\P,\P] = \R & $\so(5)$&  \checkmark\\
    \relax [H ,\B] = -\P & [H ,\P] = \B & [\B,\P] = H  &  [\B,\B]= - \R &  [\P,\P] = -\R & $\so(3,2)$& \checkmark \\\hline
    \relax & & & [\B,\B]= \B &  [\P,\P] = \B-\R && \checkmark \\
    \relax & & & [\B,\B]= \B & [\P,\P] = \R-\B &&  \checkmark\\
    \relax & & & [\B,\B] = \B & & &  \checkmark\\
    \relax & & & [\B, \B] = \P & & &  \checkmark\\
    \relax & [H ,\P] = \P & & [\B,\B] = \B & & & \\
    \relax [H ,\B] = -\P & & & [\B,\B] = \P & & & \\
    \relax [H ,\B] = \B & [H ,\P] = 2\P & & [\B,\B] = \P & &
  \end{tabular}
\end{table}

\section{$D\geq 4$}
\label{sec:dgeq-4}

We use the same abbreviated notation as in the case $D=3$ and, as in
that case, we list only those Lie brackets which do not involve the
rotational generators.  The only metric Lie algebras for $D>3$ are the
simple Lie algebras.

\begin{table}[h!]
  \centering
  \caption{Kinematical Lie algebras in $D\geq 4$}
  \label{tab:summary}
  \rowcolors{2}{blue!10}{white}
  \begin{tabular}{*{5}{>{$}l<{$}}|l|c}
    \multicolumn{5}{c|}{Nonzero Lie brackets} & \multicolumn{1}{c|}{Comments} & Metric?\\\hline
    \relax & & & & & $\s$& \\
    \relax [H,\B] = \P & & & & & $\g$& \\
    \relax [H,\B] = - \B & [H,\P] = \P & & & & $\n_-$& \\    
    \relax [H,\B] = \P & [H,\P] = - \B & & & & $\n_+$& \\
    \relax [H,\B] = \gamma \B & [H,\P] = \P & & & & $\gamma \in (-1,1]$& \\
    \relax [H,\B] = \alpha \B + \P & [H,\P] = \alpha \P - \B & & & & $\alpha > 0$& \\
    \relax [H,\B] = \B + \P & [H, \P] = \P & & & & & \\
    \relax & & [\B,\P] = H & & & $\c$& \\
    \relax [H,\B] = \P & & [\B,\P] = H & [\B,\B] = \R & & $\e$& \\
    \relax [H,\B] = - \P & & [\B,\P] = H & [\B,\B] = -\R & & $\p$& \\    
    \relax [H,\B] = \B & [H,\P] = -\P &  [\B,\P] = H + \R & & & $\so(D+1,1)$& \checkmark \\
    \relax [H,\B] = \P & [H,\P] = -\B & [\B,\P] = H &  [\B,\B]= \R &  [\P,\P] = \R & $\so(D+2)$&  \checkmark\\
    \relax [H,\B] = -\P & [H,\P] = \B & [\B,\P] = H &  [\B,\B]= - \R &  [\P,\P] = - \R & $\so(D,2)$ &  \checkmark
  \end{tabular}
\end{table}

\section{One-dimensional extensions of kinematical Lie algebras
  ($D\geq 3$)}
\label{sec:one-dimens-extens}

For $D\geq 3$, the static kinematical Lie algebra $\s$ with nonzero
brackets given by \eqref{eq:so} and \eqref{eq:kin} admits a
one-dimensional central extension $\hat\s$, with additional bracket
\begin{equation}
  [B_a, P_b] = \delta_{ab} Z\qquad (\text{or in abbreviated form }
  [\B,\P] = Z.)
\end{equation}
In \cite{JMFKinematical3D}, based on the earlier work
\cite{JMFGalilean}, we classified the deformations of $\hat\s$ for
$D=3$ and in \cite{JMFKinematicalHD} also for $D\geq 4$.  I am not
aware of any results for $D=2$, perhaps due to the fact that $\dim
H^2(\s;\RR) = 5$.  One of the central generators makes nonzero the
$[R,H]$ bracket, which would perhaps disqualify it as ``kinematical'',
since we would like to retain the identification of $R$ as a rotation.
In that case, it is the relative cohomology $H^2(\s,\r;\RR)$, with
$\r$ the one-dimensional subalgebra spanned by $R$, which one has to
calculate.  One now finds $\dim H^2(\s,\r;\RR) = 4$.  In addition, one
finds that $\dim H^2(\s,\r;\s) = 11$ and $\dim H^3(\s,\r;\s) = 29$.
We have not yet fully analysed the integrability of these
infinitesimal deformations.

Table~\ref{tab:ce-summary} lists the deformations of the universal
central extension of the static kinematical Lie algebra for $D\geq
3$.  The table is divided into three by horizontal lines.  The top
third consists of nontrivial central extensions of kinematical Lie
algebras, the middle third of trivial central extensions of
kinematical Lie algebras, and the bottom third of non-central
extensions of kinematical Lie algebras.  The metric Lie algebras here
are only the trivial extensions of the simple kinematical Lie algebras.

\begin{table}[h!]\small
  \setlength{\tabcolsep}{3pt}
  \centering
  \caption{Deformations of $\hat\s$ in $D\geq 3$}
  \label{tab:ce-summary}
  \setlength{\extrarowheight}{2pt}
  \rowcolors{2}{blue!10}{white}
  \begin{tabular}{*{6}{>{$}l<{$}}|l|c}
    \multicolumn{6}{c|}{Nonzero Lie brackets} & \multicolumn{1}{c|}{Comments}& Metric?\\\hline
  \relax [\B,\P] = Z & & & & & & $\hat\s$  & \\
  \relax [\B,\P] = Z & [H, \B] = \B & [H,\P] = -\P & & & & $\hat\n_-$  &  \\
  \relax [\B,\P] = Z & [H, \B] = \P & [H,\P] = -\B & & & & $\hat\n_+$  &  \\
    \relax [\B,\P] = Z & [H, \B] = -\P & & & & & $\hat\g$  & \\\hline
       \relax [\B,\P] = H & [H, \B] = \P & & & [\B,\B] = \R & & $\e\oplus\RR$ & \\
   \relax [\B,\P] = H & [H, \B] = - \P & & & [\B,\B] = - \R & & $\p\oplus\RR$ & \\
  \relax [\B,\P] = H + \R & [H, \B] = \B & [H, \P] = -\P & & & & $\so(D+1,1) \oplus \RR$ &  \checkmark\\
  \relax [\B,\P] = H & [H,\B] = \P & [H, \P] = - \B & & [\B,\B] = \R & [\P, \P] = \R & $\so(D+2)\oplus \RR$ &  \checkmark\\
  \relax [\B,\P] = H & [H,\B] = -\P & [H, \P] = \B & & [\B,\B] = -\R & [\P, \P] = - \R & $\so(D,2)\oplus \RR$  &  \checkmark\\\hline
   \relax [\B,\P] = Z & [H, \B] = \gamma\B & [H,\P] = \P & [H,Z] = (\gamma+1) Z& & & $\gamma \in (-1,1]$  & \\
   \relax [\B,\P] = Z & [H, \B] = \B & [H, \P] = \B + \P & [H,Z] = 2 Z & & &  & \\
   \relax [\B,\P] = Z & [H, \B] = \alpha \B + \P & [H,\P] = -\B + \alpha \P & [H,Z] = 2\alpha Z& & & $\alpha > 0$  & \\
   \relax [\B,\P] = Z & [Z,\B] = \P & [H,\P] = \P & [H, Z] = Z & [\B,\B] = \R & & $\co(D+1)\ltimes \RR^{D+1}$ & \\
   \relax [\B,\P] = Z & [Z,\B] = - \P & [H,\P] = \P & [H, Z] = Z & [\B,\B] = - \R & & $\co(D,1) \ltimes \RR^{D,1}$ & \\
  \end{tabular}
\end{table}

\section*{Acknowledgments}
\label{sec:acknowledgments}

This research is partially supported by the grant ST/L000458/1
``Particle Theory at the Higgs Centre'' from the UK Science and
Technology Facilities Council.


\providecommand{\href}[2]{#2}\begingroup\raggedright\endgroup

\end{document}